\documentclass[aps,pre,
               twocolumn,groupedaddress,
               showpacs,
               amsmath,amssymb,amsfonts,
               graphicx,floatfix]{revtex4}

\usepackage{graphicx}                            
\usepackage{dcolumn}                             
\usepackage{bm}                                  

\makeindex

\begin{document}

\title{An analytical representation for the simplest of the diffusions}

\author{E.J. Nunes-Pereira}
\affiliation{Department of Physics and Center of Physics, University of Minho, 4710-057 Braga, Portugal}
\email{epereira@fisica.uminho.pt}

\date{\today}

\begin{abstract}
  An analytical representation for the spatial and temporal dynamics of the simplest of the
  diffusions -- Brownian diffusion in an homogeneous slab geometry, with radial symmetry --
  is presented. This representation is useful since it describes the time-resolved
  (as well as stationary) radial profiles, for point-like external excitation, which are
  more important in practical experimental situations than the case of
  plane-wave external excitation. The analytical representation can be used,
  under linear system response conditions, to obtain the full dynamics for any spatial
  and temporal profiles of initial perturbation of the system. Its main value is the quantitative
  accounting of absorption in the spatial distributions. This can contribute to obtain unambiguous
  conclusions in reports of Anderson localization of classical waves in three dimensions.
\end{abstract}

\pacs{05.40.Fb, 11.80.La, 05.60.Cd, 42.25.Bs, 42.25.Dd}

\maketitle


\section{Introduction}

Diffusion, the propagation of a perturbation in space and time, is ubiquitous. It is so
ever present that the qualitative concept is used even outside the context in which it was
originally defined, like in the case of spectral diffusion or the diffusion of alleles in a
population in population genetics. The contemporary approach is
to consider diffusion, even in the strict sense of spatio-temporal spreadening of an initial
perturbation, in a broader context: one then have the classical Brownian diffusion,
the first diffusion to have a quantitative model for its description, as a particular case
of a whole possible menu for diffusion processes. Nevertheless, classical diffusion
remains the canonical simple model against which all other diffusions are gauged against.
As a result, diffusions can be labelled as diffusion (per si, the Brownian case) or
anomalous diffusion~\cite{Klages2008}. The anomalous case, rather than being exceptional, provides
the current state-of-the-art framework to approach diffusion processes. It is
subdivided into subdiffusion~(slower than the standard) or superdiffusion~(faster).
One of the alternative quantitative descriptions of anomalous diffusion uses fractional order
derivatives and it is therefore a generalization of the classical diffusion equation, although
one should state instead that the classical diffusion equation is the special case or
integer order derivatives~\cite{Metzler2004,Klages2008}. However, no matter our current
understanding of the plethora of diffusion processes, the classical,
Brownian-Einstein case remains our starting point.

If one uses a stochastic description of diffusion processes, one captures the dynamics in the
spatial and temporal probability distribution functions, the ones describing the conditional
evolution of the system, given past conditions. If the second moment of the spatial
distribution and the first moment of its temporal counterpart are finite, one can use
the classical diffusion equation~\cite{Metzler2004}. This will give a solution, valid asymptotically.
Every microscopic detail can be encapsulated in a meso or
macroscopic description using an effective diffusion coefficient. This coefficient will describe
the ensemble, in time and space scales large enough when compared to each individual
microscopic interaction. This is the physical insight which is the translation of the
more abstract concept of the central limit theorem. In the classical diffusion
case, one can encapsulate the microscopics into the mean free path, which defines a typical
scale for the interaction. The physical insight continues to build upon this: it is both a
typical scale as well as a distance which can be used to substitute the whole range of distances
for this single one. It is this insight which is used below to substitute the whole distances
for the initial perturbation, for a single individual distance corresponding to the mean free path.
There is life after a mean free path, though. In multiple scattering conditions,
after tens to thousands of mean free paths, one aims to obtain the description of the ensemble,
a cumulative response due to a large number of (possibly uncorrelated) individual
events. It is in this context that the diffusion equation is most useful and it is for
this well developed multiple scattering regime that the contribution of this manuscript is adequate.

The transport of waves is a multiple scattering regime is both well known and
very important, from the point of view of applications. The simplest of the diffusions
in the title refers to both the simplest case of diffusion-like type of interactions, the one
amenable to description by a diffusion equation, as well as its simplest realization in an
actual experimental system: a homogeneous slab finite system. In this system the diffusion
is isotropic. The historical more important case of this system is perhaps the well-known
case of plane-parallel stratified stellar atmospheres~\cite{Mihalas1978}. There is a natural
stratification in the perpendicular coordinate. It is therefore natural to use this
coordinate as the ONLY spatial relevant coordinate of the system. For an actual laboratory
setup, this means that a plane-parallel incident wave is needed, in order not to break
the symmetry. Of course, the driving force for this is that the full spatial dynamics
is amenable to a unidimensional formulation~\cite{Akkermans2006}. This is useful, to grasp the essence,
but is of difficult practical realization in a lab. The other simplifying assumption which
is common is to focus only in the stationary state description. A recent contribution was made
in which the radial profile was obtained, in steady state conditions, but not restricted
to a pseudo-unidimentional formulation~\cite{Kaiser2014}. In this context, a full solution of
the diffusion equation, describing the space and time dynamics, in multiple scattering and
in space resolved coordinates in axial symmetric conditions~(pointlike initial perturbation
and isotropic scattering in an homogeneous slab), was attempted. It is an analytical series
representation for that solution that this contribution presents. Under linear system theory
conditions, this solution can be used to obtain all the ensemble responses to any spatial
or temporal initial perturbation, by making the respective convolutions.

As a summary, the simplest diffusion in the title refers to:~(1)~multiple scattering regime,
amenable to a description by the canonical diffusion equation, (2)~isotropic interaction in an
homogeneous plane-parallel slab, maintaining axial symmetry, (3)~linear response,
and~(4)~scalar waves.

The manuscript is organized as follows:~Secs.~II and~III give an analytic representation for the
spatial and temporal dynamics, first in reflection and transmission~(II) and after inside
the slab~(III). Sec.~IV presents conclusions.


\section{Reflection and Transmission dynamics}

\begin{figure}[!htbf]
   \centering
   \includegraphics[width=\linewidth]{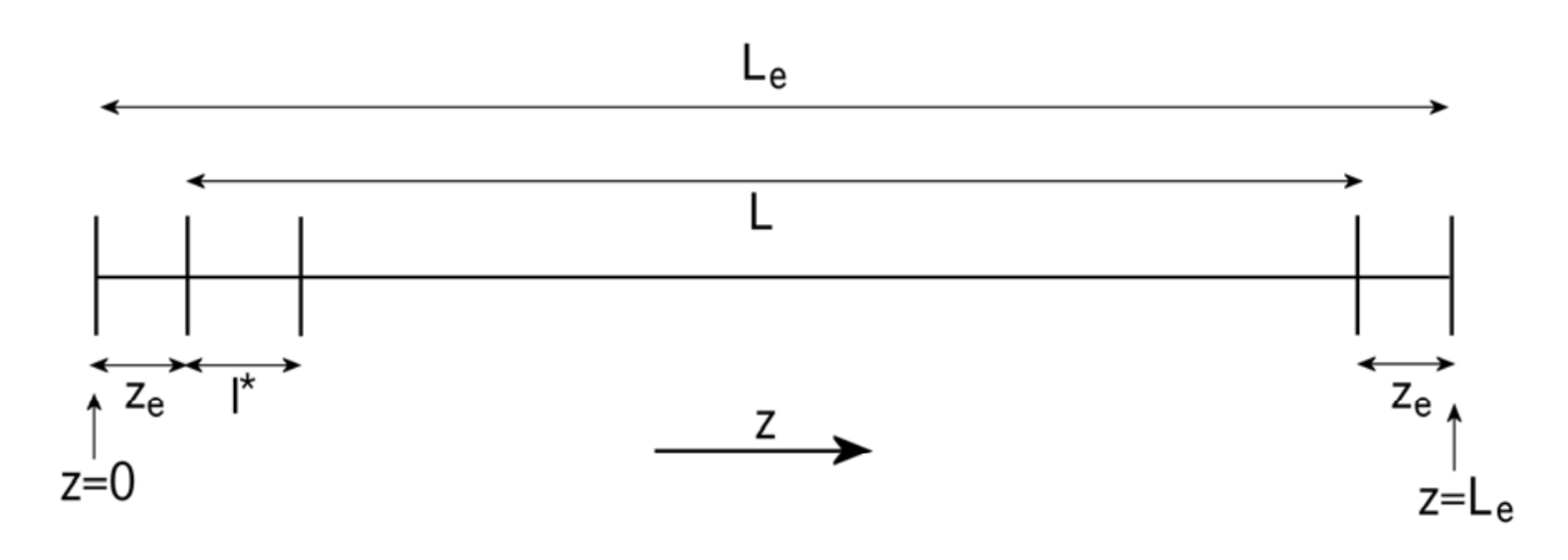}
   \caption{Notation used in derivations. $z$, $L$, $z_e$, $l^*$, and $L_e$ are the
   slab coordinate and size, an extrapolation length, the transport mean free path, and
   an effective width, respectively. Actual slab between coordinates~$z_e$, and~$L_e-z_e$.}
   \label{fig1}
\end{figure}

Fig.~\ref{fig1} shows the notation used in the derivations. The~$z$-coordinate defines the coordinate
inside slab.~$l^*$ is the transport mean free path, $L$ the slab size,
and $z_e$ a so-called extrapolation length~(this will be used in the boundary conditions for
the differential equation; its simplest value is given by~$z_e=2/3 l^*$, although more
elaborate approximations are possible, depending on refraction in the
interfaces~\cite{Akkermans2006,Durian1994}). $L_e$ is an effective width.

For axial symmetric conditions, one can group cartesian coordinates into~$z$ and~$r_\perp$,
the perpendicular distance to the symmetry axis. The conjugate coordinates for time and
space will be~$s$ and~$k_\perp$~(in this last case, conjugate to the radial distance only).

The diffusion equation describing the propagation of the intensity distribution is:

\begin{equation}
  \frac{\partial}{\partial t} I =D_{0}\frac{\partial^{2}}{\partial z^{2}} I + D_{0}\nabla_{\bot}^{2} I -\alpha I +S \mbox{,}
\end{equation}

where the symmetry is emphasized by the explicit grouping of spatial coordinates.~$I$ is the
intensity distribution, and~$D_0$, $\alpha$ and~$S$ the diffusion coefficient, absorption coefficient
and source term. The operator~$\nabla_{\bot}^{2}$ refers only to the coordinates defining
the perpendicular to the symmetry axis.

For easy reference,~$ I \equiv I \left(r_{\bot},z,t\right)$
and~$ I ^{LF}\equiv I ^{LF,2D}\left(k_{\bot},z,s\right)$.
$I \left(r_{\bot},z,t\right)$ signals axial symmetry.
$I ^{LF,2D}$ is the Laplace-Fourier transform, with 2D emphasizing that a~2D spatial Fourier
transform is implied.

The source term is now
written~$S\left(r_{\bot},z,t\right)=\delta\left(x\right)\delta\left(y\right)\delta\left(z-\left(l^{*}+z_{e}\right)\right)\delta\left(t\right)
$~(unit production plus the diffusion approximation of the whole of the external initial perturbation
being substituted for a single delta, a transport mean free path away from the sample enclosure wall;
all observables will come out normalized relative to the overall strength of the external perturbation).

In Fourier-Laplace one is left with~\cite{footnote1}:

\begin{equation}
  \left[\frac{\partial^{2}}{\partial z^{2}}-\left(k_{\bot}^{2}+\frac{s+\alpha}{D_{0}}\right)\right] I ^{LF}=-\frac{1}{\pi}\frac{1}{D_{0}}\delta\left(z-\left(l^{*}+z_{e}\right)\right) \mbox{.}
\end{equation}

Using standard Dirichelet boundary conditions~\cite{Zhu1991,Haskell1994} and the extrapolation
length, $\left\{ I\left(r_{\bot},z=0,t\right)=0,\, I\left(r_{\bot},z=L_{e},t\right)=0\right\}$
~\cite{Akkermans2006}, one obtains the solution:

\begin{eqnarray}
  \label{InsideLF1}
  I ^{LF} & = & \frac{1}{\pi}\frac{1}{D_{0}}\frac{1}{\sqrt{A}}\frac{\sinh\sqrt{A}\left(l^{*}+z_{e}\right)\sinh\sqrt{A}\left(L_{e}-z\right)}{\sinh\sqrt{A}L_{e}} \nonumber \\
  & & \; \mathrm{if}\, z<l^{*}+z_{e} \mbox{,} \\
  \label{InsideLF2}
  I ^{LF} & = & \frac{1}{\pi}\frac{1}{D_{0}}\frac{1}{\sqrt{A}}\frac{\sinh\sqrt{A}\left(L_{e}-\left(l^{*}+z_{e}\right)\right)\sinh\sqrt{A}z}{\sinh\sqrt{A}L_{e}} \nonumber \\
  & & \; \mathrm{if}\, z>l^{*}+z_{e} \mbox{,}
\end{eqnarray}

with~$A\equiv k_{\bot}^{2}+\frac{s+\alpha}{D_{0}}$.

The goal is to obtain quantities directly amenable to measurement, in simple setups.
The sought experimental observables are therefore the fluxes~($\Phi$) in the exit planes,
imaged in reflection~($R$) or transmission~($T$):

\begin{eqnarray}
  \Phi & \equiv & -D_{0}\frac{\partial I}{\partial z} \mbox{,} \\
  R & \equiv & -\Phi\left(z=z_{e},t\right) \mbox{,} \\
  T & \equiv & \Phi\left(z=L_{e}-z_{e},t\right) \mbox{.}
\end{eqnarray}

One is left with

\begin{eqnarray}
  \label{RLF}
  \Phi_{R}^{LF} & = & -\frac{1}{\pi}\frac{\sinh\sqrt{A}\left(l^{*}+z_{e}\right)\cosh\sqrt{A}\left(L_{e}-z_{e}\right)}{\sinh\sqrt{A}L_{e}} \mbox{,} \\
  \label{TLF}
  \Phi_{T}^{LF} & = & \frac{1}{\pi}\frac{\sinh\sqrt{A}\left(L_{e}-\left(l^{*}+z_{e}\right)\right)\cosh\sqrt{A}\left(L_{e}-z_{e}\right)}{\sinh\sqrt{A}L_{e}} \mbox{,}
\end{eqnarray}

whose Fourier-Lapace inversion gives:

\begin{eqnarray}
  \label{T}
  \lefteqn{T\left(r,t\right)=\frac{1}{2}\frac{1}{L_{e}^{2}}\frac{e^{-\frac{1}{4}\frac{r^{2}}{D_{0}t}}\, e^{-\alpha t}}{D_{0}t} \overset{+\infty}{\underset{n=1}{\sum}}\left(-1\right)^{n+1} \times } \nonumber \\
  & & \times n\,\cos\left(n\pi\frac{z_{e}}{L_{e}}\right)\,\sin\left(n\pi\frac{l^{*}+z_{e}}{L_{e}}\right)\, e^{-n^{2}\pi^{2}\frac{D_{0}t}{L_{e}^{2}}} \mbox{,}
\end{eqnarray}

for the transmission time resolved profile. For the reflection a similar expression is obtained,
the only difference being that the term~$\left(-1\right)^{n+1}$ is absent. PLEASE NOTE that in
all of the remaining manuscript it is shown only the Transmission results. The Reflection counterparts
are obtained simply by omitting this term. This is done in order not to overload the manuscript.

The last equation is obtained by Laplace inverse transform in the complex plane~(Bromwich integral
with poles at~$\sqrt{A}L_{e}=in\pi$) and Fourier inverse transform which, for axially
symmetric functions, takes the simple
form~$f^{2D}\left(r\right)=\int_{0}^{+\infty}F^{2D}\left(k\right)J_{0}\left(kr\right)k\, dk$,
using the zero order Bessel function~\cite{Baddour2009,Baddour2011}.

Three further important observables are total intensity, integrated in space,
~$T^{TOTAL}\left(t\right)=\int_{0}^{+\infty}\int_{0}^{2\pi} T\left(r,t\right) r\, d\theta dr$,
the steady state radial profiles,
~$T^{SS}\left(r\right)=\int_{0}^{+\infty}T\left(r,t\right)\, dt$, and the total signal
also in steady state, obtained by spatial integrating the radial
profiles~$T^{SS,TOTAL}=\int_{0}^{+\infty}\int_{0}^{2\pi}T^{SS}\left(r\right)r\, d\theta dr$:

\begin{eqnarray}
  \label{TTOTAL}
  \lefteqn{T^{TOTAL}\left(t\right)=\frac{2\pi}{L_{e}^{2}}e^{-\alpha t} \overset{+\infty}{\underset{n=1}{\sum}}\left(-1\right)^{n+1} \times } \nonumber \\
  & & \times n\,\cos\left(n\pi\frac{z_{e}}{L_{e}}\right)\,\sin\left(n\pi\frac{l^{*}+z_{e}}{L_{e}}\right)\, e^{-n^{2}\pi^{2}\frac{D_{0}t}{L_{e}^{2}}} \mbox{,} \\
  \label{TSS}
  \lefteqn{ T^{SS}\left(r\right) = \frac{1}{L_{e}^{2}}\overset{+\infty}{\underset{n=1}{\sum}}\left(-1\right)^{n+1} \times } \nonumber \\
  & & \times n\,\cos\left(n\pi\frac{z_{e}}{L_{e}}\right)\,\sin\left(n\pi\frac{l^{*}+z_{e}}{L_{e}}\right)\, \times \nonumber \\
  & & \times K_{0}\left(\sqrt{\frac{n^{2}\pi^{2}}{L_{e}^{2}}+\frac{\alpha}{D_{0}}}r\right) \mbox{,} \\
  \label{TSSTOTAL}
  \lefteqn{ T^{SS,TOTAL} = \frac{2}{\pi}\overset{+\infty}{\underset{n=1}{\sum}}\left(-1\right)^{n+1} \times } \nonumber \\
  & & \times \frac{1}{n\left(1+\frac{\alpha L_{e}^{2}}{D_{0}n^{2}\pi^{2}}\right)}\,\cos\left(n\pi\frac{z_{e}}{L_{e}}\right)\,\sin\left(n\pi\frac{l^{*}+z_{e}}{L_{e}}\right) \mbox{,}
\end{eqnarray}

where~$K_{0}$ is the modified Bessel function of the second kind.

Eqs.~(\ref{T}-\ref{TSSTOTAL}) constitute the main contribution of this manuscript. Compare also the
time resolved total~(not spatial resolved) fluxes in refs.~\cite{Maret2000} and~\cite{Johnson2003}.

Please note that, without absorption~($\alpha=0$)

\begin{eqnarray}
  \lefteqn{ R^{SS,TOTAL}+T^{SS,TOTAL}= } \nonumber \\
  & = & \frac{4}{\pi}\overset{+\infty}{\underset{n=0}{\sum}}\frac{1}{2n+1}\, \cos\left(\left(2n+1\right)\pi\frac{z_{e}}{L_{e}}\right)\, \times \nonumber \\
  & & \times \sin\left(\left(2n+1\right)\pi\frac{l^{*}+z_{e}}{L_{e}}\right)=1 \mbox{,}
\end{eqnarray}

which express energy conservation~(and is therefore a simple consistency check).

Eqs.~(\ref{T}-\ref{TSSTOTAL}) are best recast in dimensionless coordinates. The characteristic scales are:

\begin{eqnarray}
  r' & = & \frac{r}{L_{e}} \mbox{,} \\
  t' & = & \frac{D_{0}t}{L_{e}^{2}} \mbox{,} \\
  \alpha' & = & \frac{L_{e}^{2}\alpha}{D_{0}} \mbox{,}
\end{eqnarray}

and the previous results become:

\begin{eqnarray}
  \label{Tless}
  \lefteqn{ T\left(r',t'\right) \equiv \frac{T\left(r,t\right)}{\frac{1}{2 D_{0} L_{e}}} = \frac{e^{-\frac{1}{4}\frac{r'^{2}}{t'}}\, e^{-\alpha't'}}{t'}\overset{+\infty}{\underset{n=1}{\sum}}\left(-1\right)^{n+1} \times } \nonumber \\
  & & n\,\cos\left(n\pi\frac{z_{e}}{L_{e}}\right)\,\sin\left(n\pi\frac{l^{*}+z_{e}}{L_{e}}\right)\, e^{-n^{2}\pi^{2}t'} \mbox{,} \\
  \label{TTOTALless}
  \lefteqn{ T^{TOTAL}\left(t'\right) \equiv \frac{T^{TOTAL}\left(t\right)}{\frac{2\pi}{D_{0}}} = e^{-\alpha't'}\overset{+\infty}{\underset{n=1}{\sum}}\left(-1\right)^{n+1} \times } \nonumber \\
  & & n\,\cos\left(n\pi\frac{z_{e}}{L_{e}}\right)\,\sin\left(n\pi\frac{l^{*}+z_{e}}{L_{e}}\right)\, e^{-n^{2}\pi^{2}t'} \mbox{,} \\
  \label{TSSless}
  \lefteqn{ T^{SS}\left(r'\right) \equiv \frac{T^{SS}\left(r\right)}{\frac{1}{L_{e}}} = \overset{+\infty}{\underset{n=1}{\sum}}\left(-1\right)^{n+1} \times } \nonumber \\
  & & n\,\cos\left(n\pi\frac{z_{e}}{L_{e}}\right)\,\sin\left(n\pi\frac{l^{*}+z_{e}}{L_{e}}\right)\, \times \nonumber \\
  & & K_{0}\left(\sqrt{n^{2}\pi^{2}+\alpha'}r'\right) \mbox{,} \\
  \label{TSSTOTALless}
  \lefteqn{ T^{SS,TOTAL} = \frac{2}{\pi}\overset{+\infty}{\underset{n=1}{\sum}}\left(-1\right)^{n+1} \times } \nonumber \\
  & & \frac{1}{n\left(1+\frac{\alpha'}{n^{2}\pi^{2}}\right)}\,\cos\left(n\pi\frac{z_{e}}{L_{e}}\right)\,\sin\left(n\pi\frac{l^{*}+z_{e}}{L_{e}}\right) \mbox{.}
\end{eqnarray}

Two aspects are worth mentioning. The first is the asymptotic behaviour, in space and time.
This is obtained in the previous equations making~$n=1$, which is the slowest decaying term.
The asymptotics are:

\begin{eqnarray}
  T\left(r',t'\right) & \propto & \frac{e^{-\frac{1}{4}\frac{r'^{2}}{t'}}\, e^{-\left(\pi^{2}+\alpha'\right)t'}}{t'} \mbox{,} \\
  \label{asymptotic}
  T^{TOTAL}\left(t'\right) & \propto & e^{-\left(\pi^{2}+\alpha'\right)t'} \mbox{,} \\
  T^{SS}\left(r'\right) & \propto & K_{0}\left(\sqrt{\pi^{2}+\alpha'}r'\right) \mbox{.}
\end{eqnarray}

The time-resolved distributions are Gaussian~(with variance linear with time).
The total signal, integrated in space,
is exponential with~$t'$. And the steady state radial distributions are quasi-exponential with the
radial distance~$r'$. This last statement is due to the facts~\cite{Kaiser2014,Duffy2001}:
~(1)~$K_{\alpha}\left(x\right)=\frac{\pi}{2} i^{\alpha+1} H_{\alpha}^{(1)}\left(i x\right)$,
for~$-\pi < arg(x) \leq \pi$, and
(2)~$K_{0}\left(\sqrt{\pi^{2}+\alpha'}r'\right) \propto r'^{-1/2} exp\left(-\sqrt{\pi^{2}+\alpha'}r'\right)$.
The~$r'^{-1/2}$ has a much weaker dependence on the distance than the exponential, and a
approximate exponential dependence follows.

The second aspect is the mean displacement, as can be judged from the radial profiles in
reflection or transmission. It is well known that Brownian diffusion gives a
mean square displacement that is linear with time in two dimensions. In fact, the
time dependence of this mean square displacement is the most important parameter
distinguishing classical from anomalous diffusion. It is therefore
tempting to estimate the mean radial \emph{displacement}, \emph{relative to the symmetry axis}, in the
imaged fluxes.
The result is~$\left\langle r'^{2}\right\rangle \propto t'$,
showing that, the physical signature of classical diffusion survives in the imaged fluxes for
the three dimensional slab~(see also~\cite{Akkermans2006,Maret2013,VanTiggelen2008}).

\begin{figure}[!htbf]
   \centering
   \includegraphics[width=\linewidth]{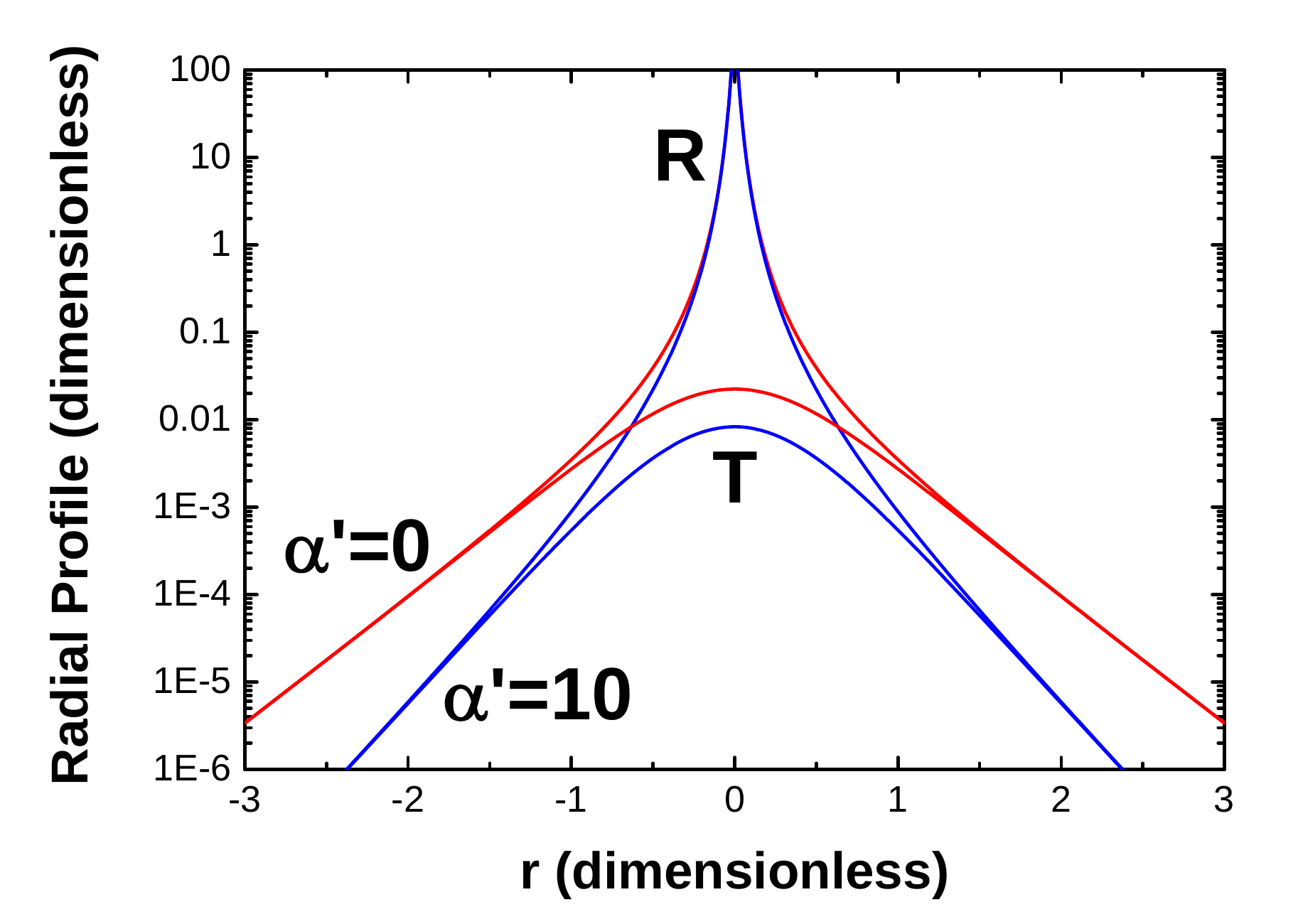}
   \caption{(Color online)~Steady state radial profiles, in transmission~(T) and reflection~(R), using
   eq.~(\ref{TSSless})~($2000$ terms, $l^*=1$, $L_e=50$, $z_e=2/3 l^*$). $\alpha'$ is a normalized
   absorption coefficient.}
   \label{fig2}
\end{figure}

\begin{figure}[!htbf]
   \centering
   \includegraphics[width=\linewidth]{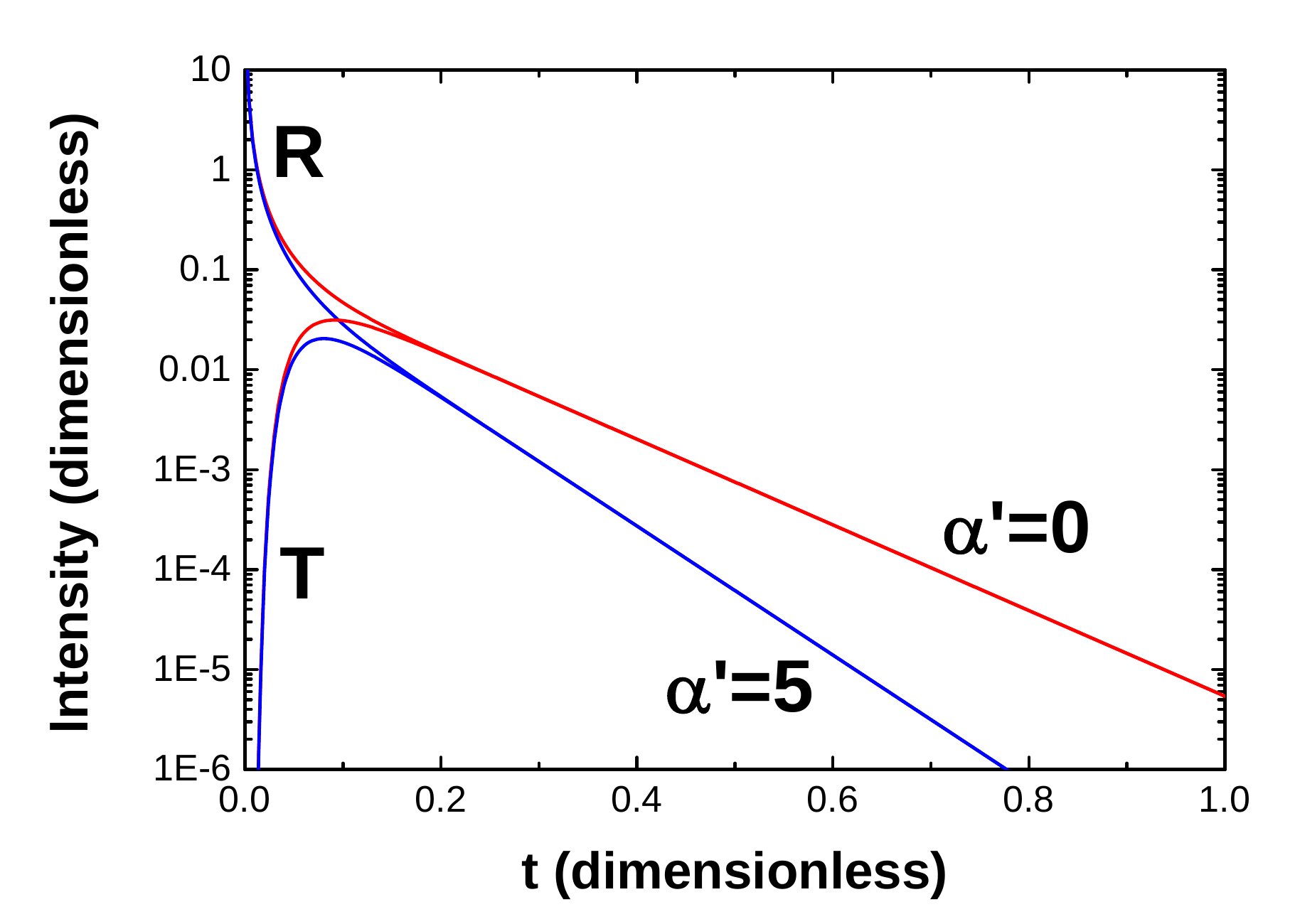}
   \caption{(Color online)~Time resolved total intensity in transmission~(T) and reflection~(R), using
   eq.~(\ref{TSSless})~($2000$ terms, $l^*=1$, $L_e=50$, $z_e=2/3 l^*$). $\alpha'$ is a normalized
   absorption coefficient.}
   \label{fig3}
\end{figure}

\begin{figure}[!htbf]
   \centering
   \includegraphics[width=\linewidth]{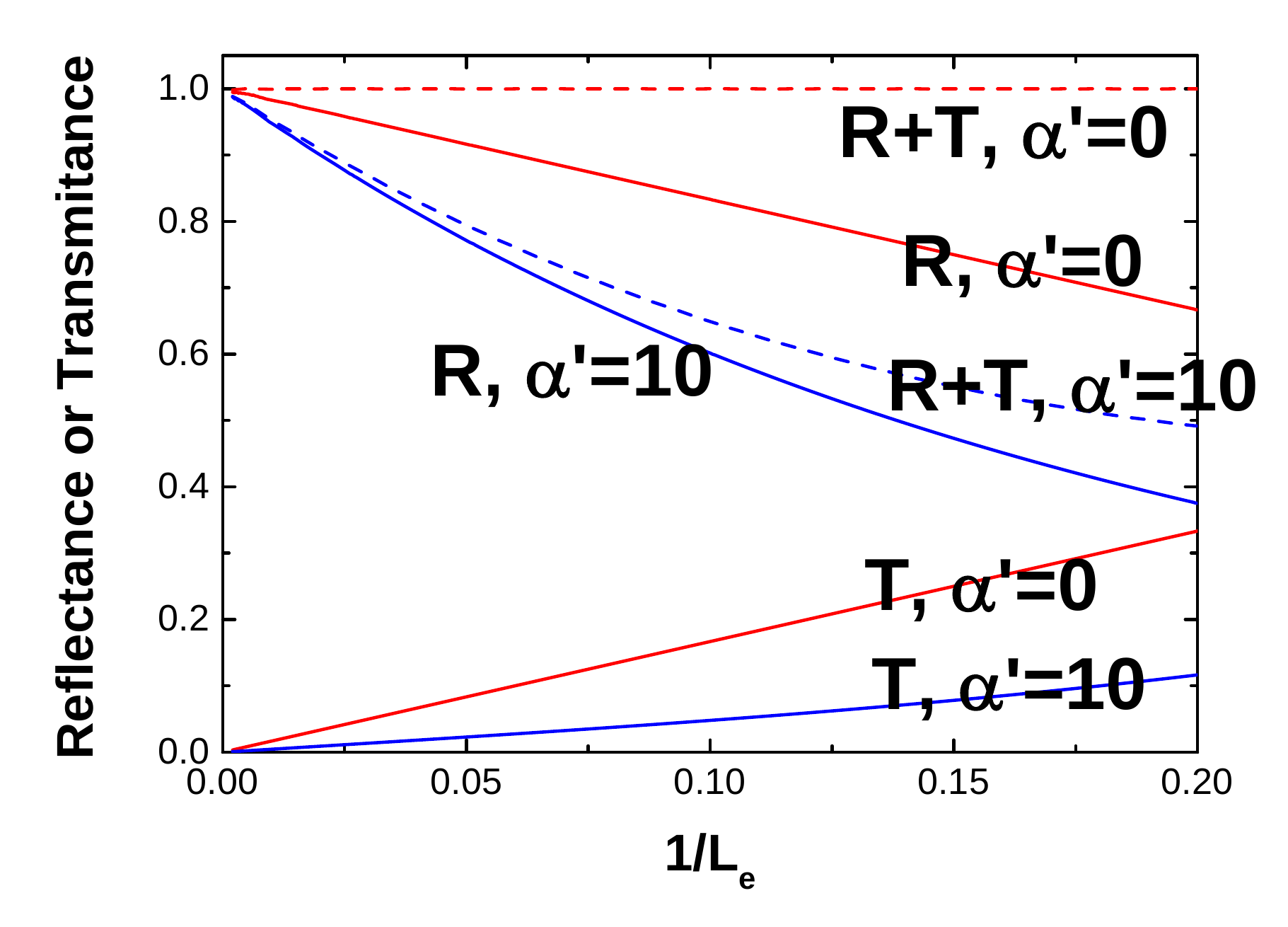}
   \caption{(Color online)~Trnasmitance~(T), reflectance~(R) and sum~(T+R), as a function of inverse efective
   slab width, using eq.~(\ref{TSSTOTALless})~($2000$ terms, $l^*=1$, $z_e=2/3 l^*$). $\alpha'$ is a normalized
   absorption coefficient. Without absorption, $R+T=1$ which is a manifestation of energy conservation.}
   \label{fig4}
\end{figure}

Figs.~\ref{fig2} to~\ref{fig4} provide a graphical illustration of the results in
eqs.~(\ref{Tless}-\ref{TSSTOTALless}). Fig.~\ref{fig2} shows the radial profiles, in steady state,
illustrating the quasi-exponential radial asymptotic. This asymptotic regime was recently verified
experimentally~\cite{Kaiser2014}. Fig.~\ref{fig3} shows time resolved total transmission and reflection,
illustrating also the asymptotics of eq.~(\ref{asymptotic}). Fig.~\ref{fig4}
shows that the series in the previous equations reproduce the trend known for the
diffusion in the slab: overall transmission in steady-state in inversely proportional to
slab width, in the absence of absorption~\cite{Akkermans2006}.
It further shows that this scaling is broken, if absorption is present.

One important aspect in the derivations was that the slab was considered infinite,
in the transverse directions. The finite size could destroy the quasi-exponential
asymptotics of the radial profiles. Reference~\cite{Kaiser2014} gives empirical evidence that
this asymptotic is amenable to experimental measurement, for an actual slab
representative of simple experimental conditions~(aspect ratio of slab diameter to width
roughly~$10$).

The spatial profiles in fig.~\ref{fig2} needed special care in their evaluation near the origin. The modified Bessel function of the second kind diverge at the origin. Although it is not provided an analytical proof of the series convergence, an empirical study of the series convergence was nevertheless conducted. Using a sufficient high number of terms, it was verified in all cases the series convergence. The convergence is however very slow. The fixed number of terms in fig.~\ref{fig2} correspond to a conservative estimate, based in the smallest distance considered~(see also~\cite{Kaiser2014}).


\section{Inside of slab dynamics}

The results presented thus far are focused in the dynamics of the transmission and reflection fluxes,
since these are the ones most easily measured. The dynamics inside the slab is nevertheless of
importance. The procedures used in the inverse Laplace and Fourier transforms of
eqs.~(\ref{RLF}-\ref{TLF}), can be equally applied to obtain the dynamics inside the slab.
Eqs.~(\ref{InsideLF1}-\ref{InsideLF2}) give, first for the dynamics in direct physical variables

\begin{eqnarray}
  \lefteqn{ I\left(r,z,t\right) = \frac{1}{2\pi L_{e}}\frac{e^{-\frac{1}{4}\frac{r^{2}}{D_{0}t}}\, e^{-\alpha t}}{D_{0}t} \overset{+\infty}{\underset{n=1}{\sum}}\sin\left(n\pi\frac{l^{*}+z_{e}}{L_{e}}\right)\, \times } \nonumber \\
  & & \times \sin\left(n\pi\frac{z}{L_{e}}\right)\, e^{-n^{2}\pi^{2}\frac{D_{0}t}{L_{e}^{2}}} \mbox{,} \\
  \lefteqn{ I^{SS}\left(r,z\right) = \frac{1}{\pi D_{0}L_{e}}\overset{+\infty}{\underset{n=1}{\sum}}\sin\left(n\pi\frac{l^{*}+z_{e}}{L_{e}}\right)\, \times } \nonumber \\
  & & \times \sin\left(n\pi\frac{z}{L_{e}}\right)\, K_{0}\left(\sqrt{\frac{n^{2}\pi^{2}}{L_{e}^{2}}+\frac{\alpha}{D_{0}}}r\right) \mbox{,} \\
  \lefteqn{ I^{SS,TOTAL}\left(z\right) = \frac{2}{\pi^{2}}\frac{L_{e}}{D_{0}}\overset{+\infty}{\underset{n=1}{\sum}}\frac{1}{n^{2}\left(1+\frac{\alpha L_{e}^{2}}{D_{0}n^{2}\pi^{2}}\right)}\, \times } \nonumber \\
  & & \times \sin\left(n\pi\frac{l^{*}+z_{e}}{L_{e}}\right)\,\sin\left(n\pi\frac{z}{L_{e}}\right) \mbox{,}
\end{eqnarray}

and, then for its reformulation in dimensionless coordinates

\begin{eqnarray}
  \lefteqn{ I\left(r',z,t'\right) \equiv \frac{I\left(r,z,t\right)}{\frac{1}{2\pi D_{0}}} = \frac{e^{-\frac{1}{4}\frac{r'^{2}}{t'}}\, e^{-\alpha't'}}{t'} \times } \nonumber \\
  & & \times \overset{+\infty}{\underset{n=1}{\sum}}\sin\left(n\pi\frac{l^{*}+z_{e}}{L_{e}}\right)\,\sin\left(n\pi\frac{z}{L_{e}}\right)\, e^{-n^{2}\pi^{2}t'} \mbox{,} \\
  \lefteqn{ I^{SS}\left(r',z\right) \equiv \frac{I^{SS}\left(r,z\right)}{\frac{1}{\pi D_{0}}} = \overset{+\infty}{\underset{n=1}{\sum}}\sin\left(n\pi\frac{l^{*}+z_{e}}{L_{e}}\right)\,\sin\left(n\pi\frac{z}{L_{e}}\right)\, \times } \nonumber \\
  & & \times K_{0}\left(\sqrt{n^{2}\pi^{2}+\alpha'}r'\right) \mbox{,} \\
  \label{InsideSS}
  \lefteqn{ \frac{I^{SS,TOTAL}\left(z\right)}{\frac{2}{\pi^{2}}\frac{L_{e}}{D_{0}}} = \overset{+\infty}{\underset{n=1}{\sum}}\frac{1}{n^{2}\left(1+\frac{\alpha'}{n^{2}\pi^{2}}\right)}\, \times } \nonumber \\
  & & \times \sin\left(n\pi\frac{l^{*}+z_{e}}{L_{e}}\right)\,\sin\left(n\pi\frac{z}{L_{e}}\right) \mbox{.}
\end{eqnarray}

\begin{figure}[!htbf]
   \centering
   \includegraphics[width=\linewidth]{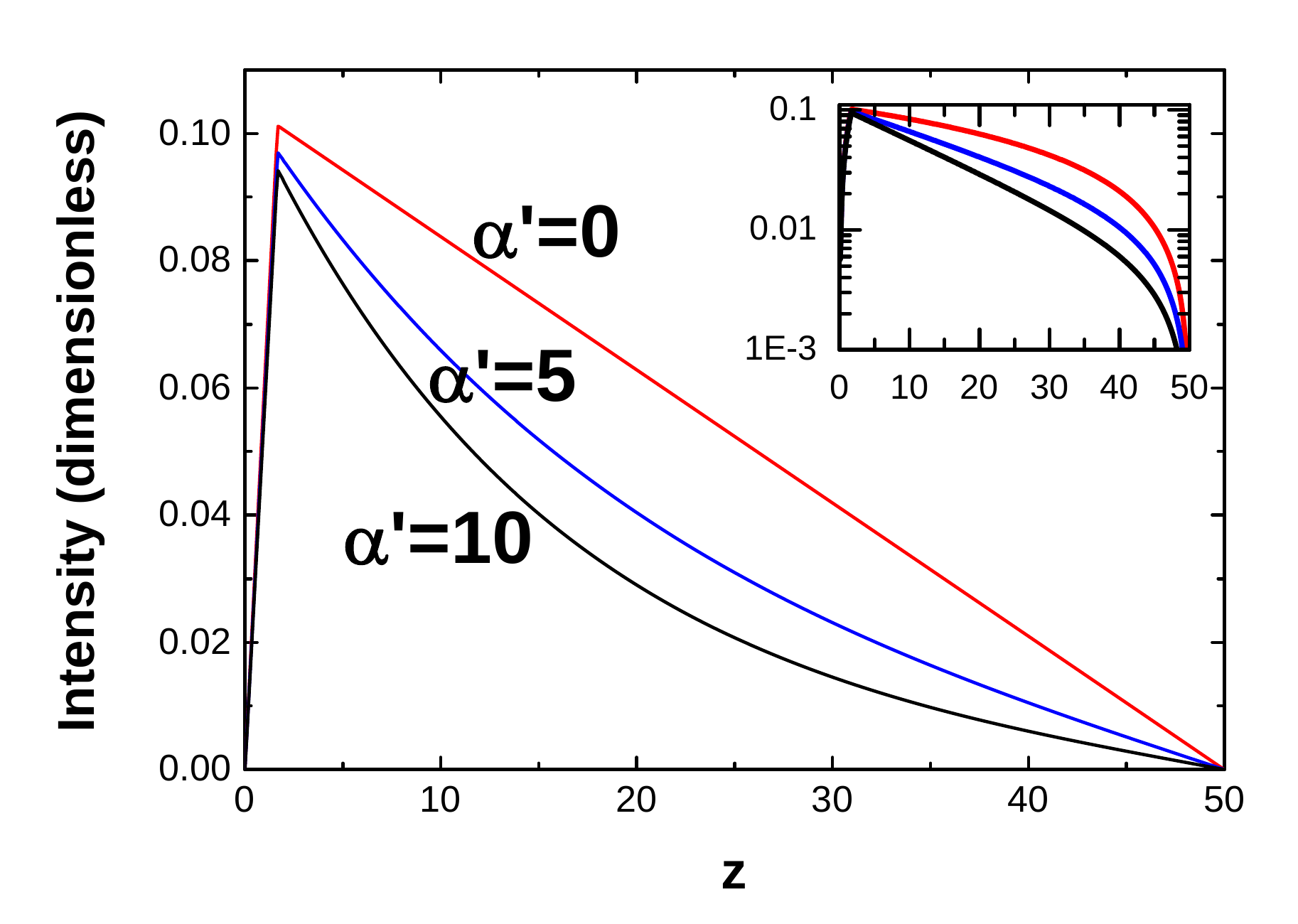}
   \caption{(Color online)~Steady state intensity distribution inside slab, using
   eq.~(\ref{InsideSS})~($2000$ terms, $l^*=1$, $L_e=50$, $z_e=2/3 l^*$). Total intensity, integrated
   in a plane perpendicular to the $z$ coordinate.  $\alpha'$ is a normalized
   absorption coefficient.
   Actual slab between coordinates~$z_e$, and~$L_e-z_e$.}
   \label{fig5}
\end{figure}

Finally, fig.~\ref{fig5} represents the steady-state intensity inside the slab, integrated over
radial distances, as a function
of spatial coordinate. This result encompasses the known trend, for a unidimensional
formulation of diffusion in absence of absorption:~the intensity is linear in the difference
to a coordinate inside the slab, situated one mean free path away from the
boundary~(this position comes from the approximation in the source term, the linear
dependence comes from diffusion itself)~\cite{Johnson2003,Maret2000}. It further quantifies to what
extend absorption destroys the linear dependence with distance. The inset gives evidence that,
at least for strong absorption, the dependence with distance can be reasonably
approximated by an exponential~(see also~\cite{Akkermans2006}).

\section{Conclusions}

This contribution presents an analytical representation for the full dynamics in the
simplest of the possible realizations of a diffusion process:~classical diffusion of scalar waves in a
fully developed multiple scattering regime, in an homogeneous plane-parallel slab. These
conditions preserve axial symmetry, which was exploited in the derivations. The analytical
series is strictly valid for a unit initial perturbation. However, under linear system response
conditions, the convolution with the spatial and temporal distributions of the
perturbation allow a solution for all other cases. Although this is conceptually trivial, even a numerical implementation of a~2D spatial convolution of two radial
symmetric functions~(initial external perturbation and system response) must
implement a proper definition of the convolution as a \emph{two-dimensional}
convolution~\cite{Baddour2011}. For other alternative approaches for taking into account
the finite dimensions of the external perturbation see refs.~\cite{Akkermans2006,Ishimaru1983}. Ref.~\cite{Baddour2011} can additionally provide a \emph{basis} for a generalization of this 
contribution to non axial symmetric situations.

It is expected that this contribution stimulates quantitative detailed comparison with
experimental results, obtained in actual setups not amenable to a realistic
unidimensional formulation of the geometry. Its main contribution is the derivation of the
spatial resolved profiles, including absorption. These can help in obtaining unambiguous conclusions
in reports of Anderson localization of classical waves in three dimensions, a subject of current
interest and debate~\cite{Maret2013,VanTiggelen2008,correspondence}.

\begin{acknowledgements}
   The author would like to acknowledge fruitful discussions with
   R.~Kaiser (CNRS, INLN), R.~Pierrat (ESPCI, CNRS), N.~Baddour (U.Ottawa), and B.P.J.~Bret (Bosch).
\end{acknowledgements}

\end{document}